\documentclass[3p,times,procedia]{elsarticle}
\usepackage{nupha_ecrc}
\usepackage[left]{lineno}
\usepackage{url}

\volume{00}

\firstpage{1}

\journalname{Nuclear Physics A}

\runauth{}

\jid{nupha}

\jnltitlelogo{Nuclear Physics A}

\usepackage{amssymb}
\usepackage[osf,sc]{mathastext}
\usepackage{upgreek}

\usepackage[figuresright]{rotating}

\begin{document}

\begin{frontmatter}



\dochead{XXVIIIth International Conference on Ultrarelativistic Nucleus-Nucleus Collisions\\ (Quark Matter 2019)}

\title{Studies of Strange and Non-Strange Beauty Productions in PbPb Collisions with the CMS Detector}


\author{Zhaozhong Shi, on behalf of the CMS Collaboration}

\address{Laboratory for Nuclear Science, Massachusetts Institute of Technology, Cambridge, MA USA 02139}

\begin{abstract}

Beauty quarks are considered as one of the best probes of the strongly interacting medium created in relativistic heavy-ion collisions because they are predominantly produced via initial hard scatterings. Our measurements of B mesons productions provide information on the diffusion of beauty quarks and the flavor dependence of in-medium energy loss. In these studies, clarifying the hadronization mechanism is crucial for understanding the transport properties of beauty quarks. We propose to show our experimental results on three sets of particles: nonprompt $J/\psi$ and $D^0$, $B^0_s$ and $B^+$, and prompt $D^+_s$ and $D^0$. Our measurements of $B^0_s$ production can shed light on the mechanisms of beauty recombination in the medium. In addition, the measurements of mesons productions containing both strange and beauty quarks can provide more information about strangeness enhancement in the quark-gluon plasma. The ratio of $B^0_s$ over $B^+$ nuclear modification factors at a nucleon-nucleon center-of-mass energy of 5.02 TeV, using fully reconstructed B mesons with the CMS detector, are presented. We will also show the nonprompt $J/\psi$ and $D^0$ from the partial B mesons decay chains as well as our precise measurement the prompt $D^+_s$ to $D^0$ ratio in pp and PbPb collisions.

\end{abstract}


\end{frontmatter}


\section{Introduction}

In high-energy heavy-ion collisions, heavy quarks, including charm and bottoms quarks, are used as probes to understand the formation and evolution of the quark-gluon plasma (QGP). Because the mass of charm and beauty quarks are on the order of a few GeV, they are predominantly created at the early stage of heavy ion collisions when hard scattering processes occur. Because they have longer decay life times than the QGP, they retain their identities and record the evolution the of QGP. Since in general they do not reach thermal equilibrium and have long thermal relaxation times, they can be used as probes to study the transport properties of the QGP. 

Similar to light quarks and gluons, in parton shower stage, heavy quarks lose significant fractions of energies as they travel through the QGP medium \cite{HQEnergyLoss}. In perturbative QCD, the coupling strength of heavy quarks to the medium is assumed to be weak. In this microscopic picture, heavy quarks scatter off the QGP constituents and undergo medium-induced gluon radiation \cite{pQCDHFQGP}. They diffuse and lose energy as they traverse through the medium. There are two energy loss mechanisms of heavy quarks in the QGP medium: collisional energy loss and radiative energy loss. The collisional energy loss is described by the momentum transfer from heavy quarks to medium constituents \cite{Collisional} in the elastic scattering processes. The gluon radiations of heavy quarks are modified with the influence of the QGP medium compared to vacuum \cite{Radiative}. In addition, the dead cone effect \cite{DeadCone} predicts the flavor dependence of parton energy loss. We expect to observe the suppression of the heavy flavor hadrons production. Moreover, heavy flavor hadrons should lose less energy than light flavor hadrons.

Because the temperature of the thermally and chemically equilibrated QGP lies above the strange quark mass, the strangeness content in the QGP is expected to be enhanced via the process $gg \rightarrow s \bar s$. Heavy quarks may hadronize along with the medium constituents via the recombination mechanism. Therefore, we expect to observe enhancement of strange heavy flavor hadrons compare to their non-strange counterparts \cite{SQEnhance} with the creation of QGP. Hence, we can use the production of strange and non-strange heavy flavor hadrons to study the QGP medium effects on hadronization and probe the internal structure of the QGP.

We use the experimental observable nuclear modification factor $R_{AA}$, which quantifies the modification on the particle spectra with the presence of the QGP medium, to decipher the energy loss mechanisms of heavy quarks and understand the QGP medium properties. To study heavy quarks hadronization mechanism, we use the cross-section ratios of strange and non-strange heavy flavor hadrons in pp and AA collisions to quantify the strangeness enhancement associated to the QGP medium. Hence, in the paper, we will present the studies of the spectra following three sets of particles: nonprompt $J/\psi$ and $D^0$ mesons, $B^0_s$ and $B^+$ mesons, as well as prompt $D_s^{+}$ and $D^0$ mesons with the CMS detector \cite{CMSDetector}.

\section{Nonprompt $J/\psi$ mesons and nonprompt $D^0$ mesons}



The nonprompt $J/\psi$ and $D^0$ both come from the B meson decay. The nonprompt $J/\psi$ mesons are reconstructed from the dimuon decay channel. We apply muon identification and kinematic selections to select the $ J/\psi$ candidates. Nonprompt $D^0$ mesons are reconstructed in the $D^0 \rightarrow K^- \pi^+$ channel. Boosted decision trees (BDTs), using the daughter track variables as inputs, are implemented to optimally select $D^0$ candidates. No hadronic particle identification is used to reconstruct $D^0$ mesons.



\begin{figure}[hbtp]
\begin{center}
\includegraphics[width=0.32\textwidth]{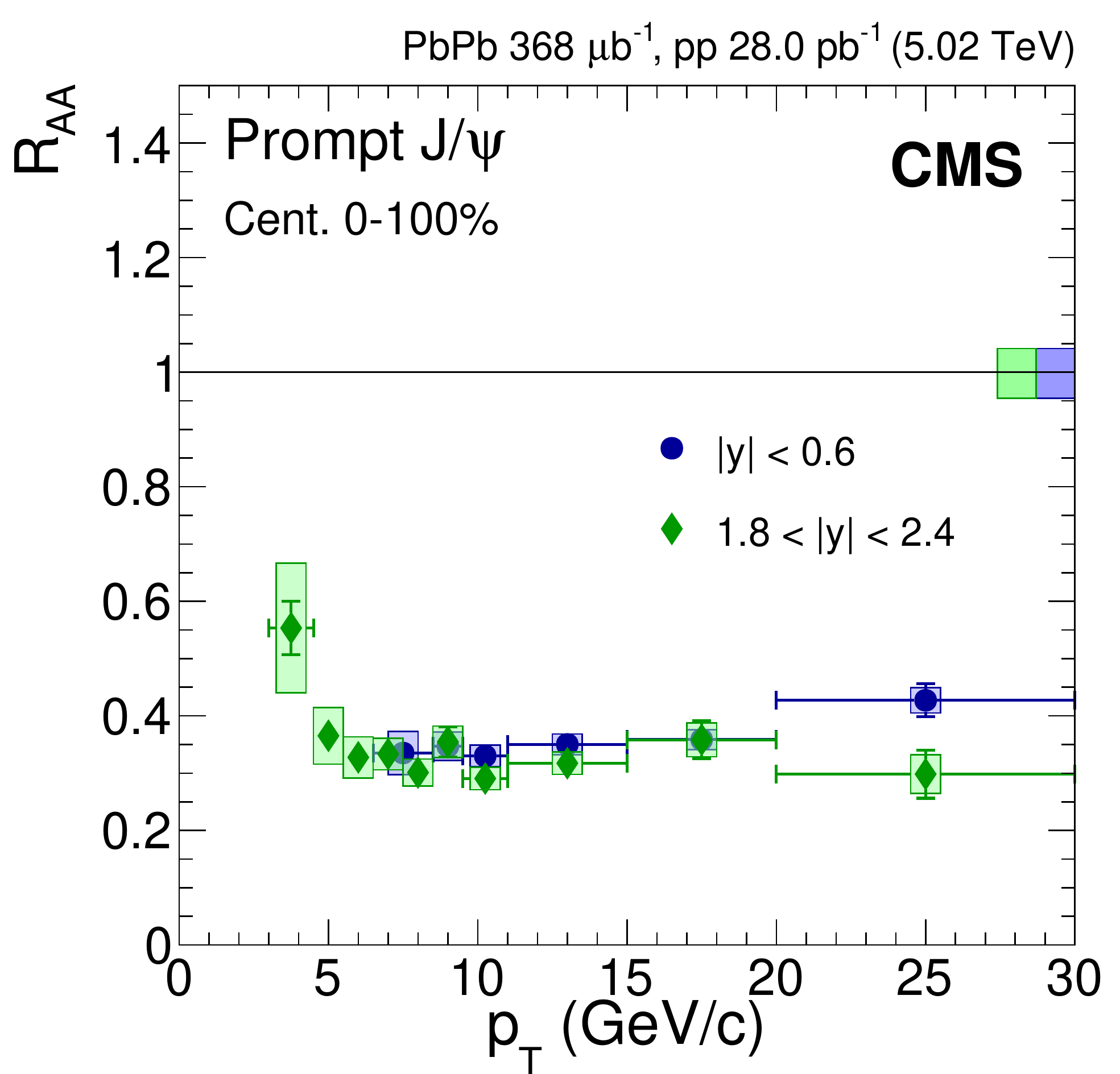}
\includegraphics[width=0.32\textwidth]{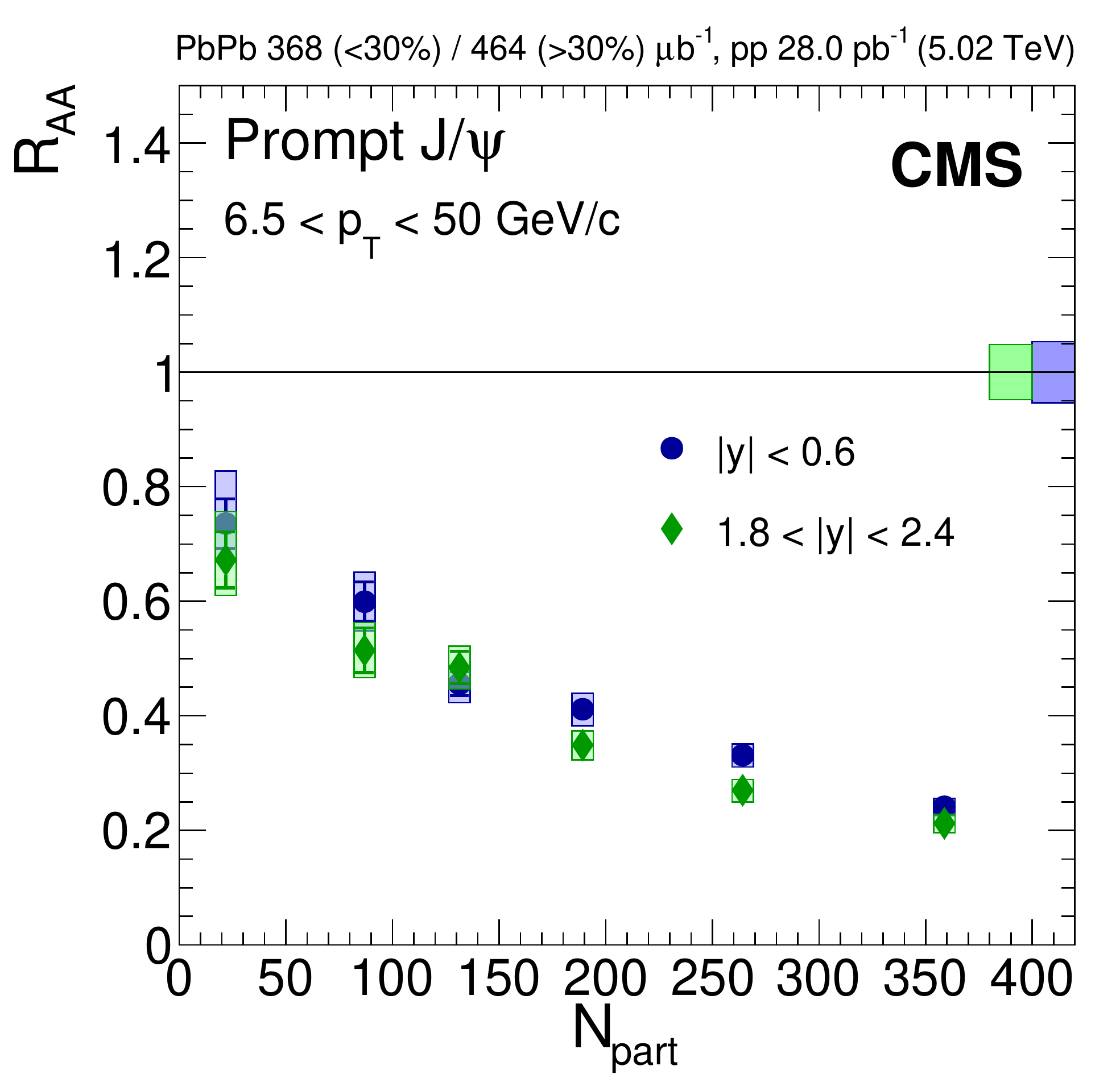}
\caption{The nonprompt $J/\psi$ $R_{AA}$ as a function of $p_T$ (left) and as a function of $N_{part}$ (right) in pp and PbPb collisions at $\sqrt{s_{NN}}$ = 5.02 TeV \cite{NPJPsi}.}
\label{fig:NPJPsiRAA}
\end{center}
\end{figure}

After extracting the nonprompt component and correcting for efficiency and luminosity, we get the nonprompt $J/\psi$ $R_{AA}$ as a function of $p_T$ and $N_{part}$ in pp and PbPb collisions at $\sqrt{s_{NN}}$ = 5.02 TeV in Fig~\ref{fig:NPJPsiRAA}. $N_{part}$ stands for the number of participants nucleons in the collisions.
 
From Fig.~\ref{fig:NPJPsiRAA}, a significant suppression of nonprompt $J/\psi$ in PbPb collision compared to pp collision and a larger suppression of the nonprompt $J/\psi$ in more central collisions is observed. This suggests that bottom quarks lose energy in QGP before weakly decaying to charm quarks. In addition, there is no significance $p_T$ dependence of nonprompt $J/\psi$ at $p_T >$ 5 GeV/c.





\begin{figure}[hbtp]
\begin{center}
\includegraphics[width=0.35\textwidth]{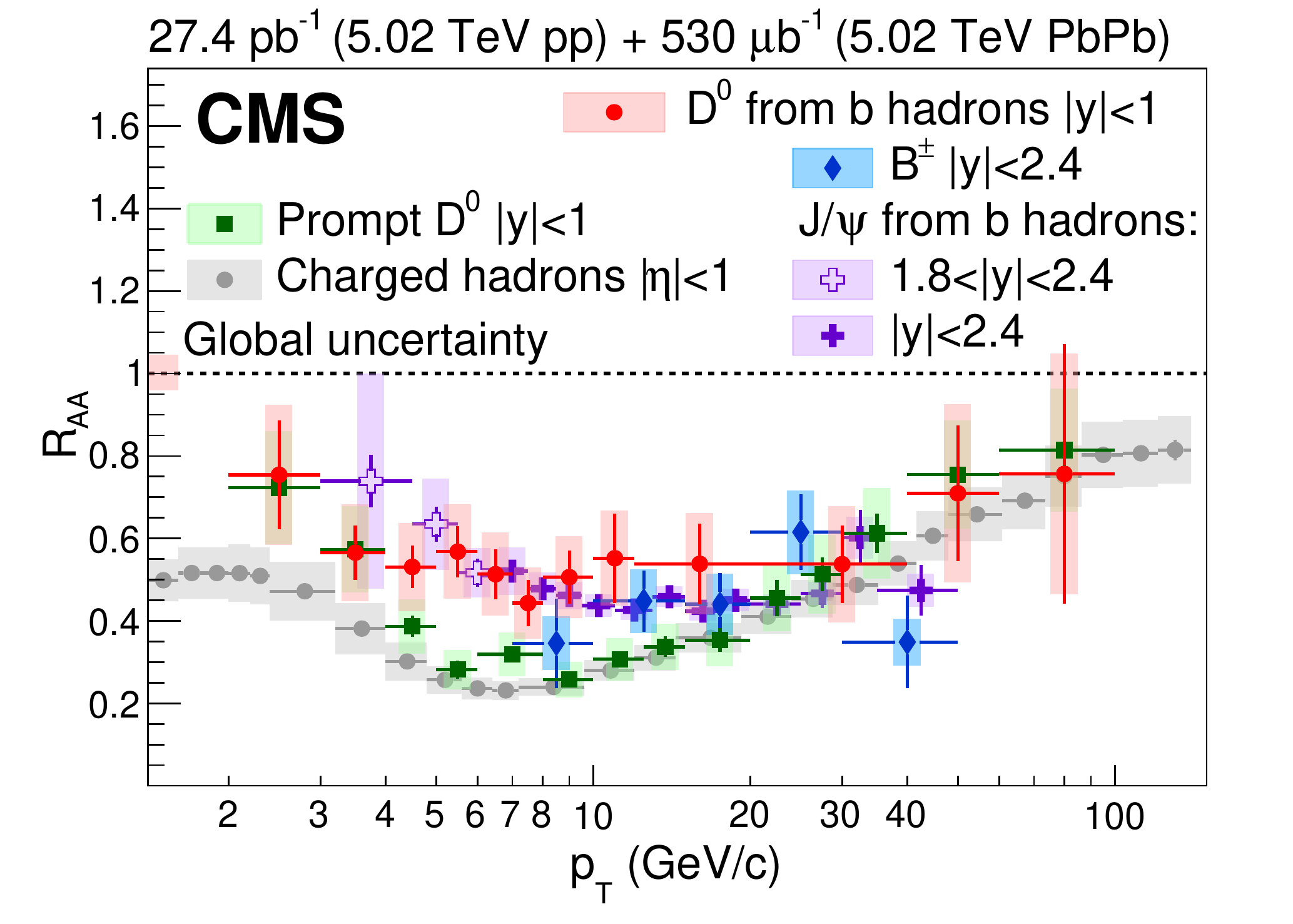}
\includegraphics[width=0.375\textwidth]{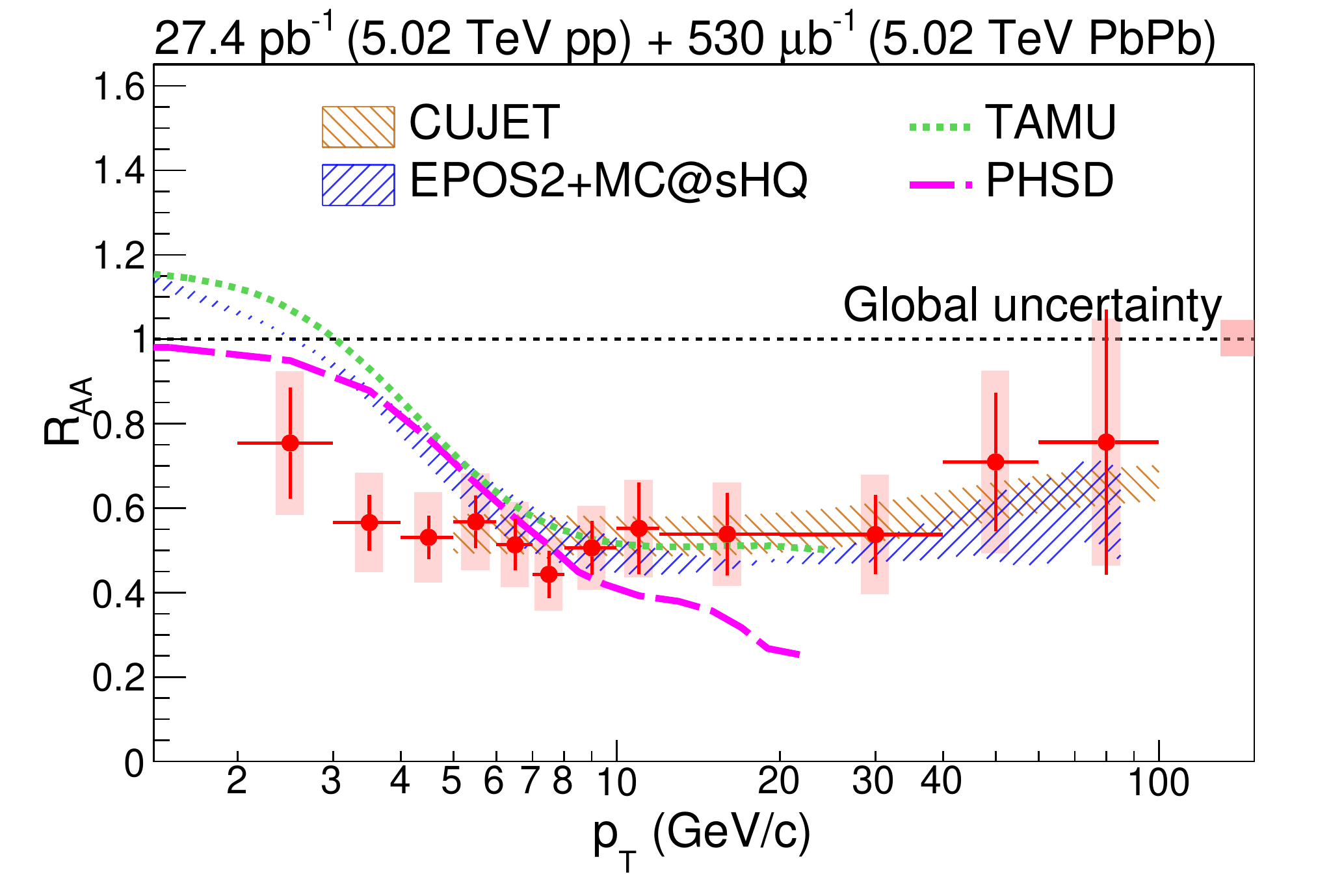}
\caption{The nonprompt $D^0$ $R_{AA}$ as a function of $p_T$ at 5.02 TeV in pp and PbPb collisions \cite{NPD0RAACite} are presented. We compare the $R_{AA}$ as a function of $p_T$ of prompt $D^0$ \cite{PromptD0}, charged hadrons \cite{chagredhadrons}, $B^\pm$\cite{BPlusPaper}, and nonprompt $J/\psi$ for $|y| < 2.4$ and $1.8 < |y| < 2.4$ \cite{NPJPsi}. The data are also compared with the calculations from CUJET \cite{CUJET}, EPOS2+MC@sHQ \cite{EPOS}, TAMU \cite{D0TAMU}, and PHSD \cite{PHSD}.}
\label{fig:NPD0RAA}
\end{center}
\end{figure}

Our results of nonprompt $D^0$ $R_{AA}$ as a function $p_T$ and comparison with other hadrons and theoretical calculations are summarized on Fig. \ref{fig:NPD0RAA}.

In general, the production of nonprompt $D^0$ is suppressed between 2 and 100 GeV/c in PbPb compared to pp. Comparing nonprompt $D^0$ $R_{AA}$ with other hadron species, we can see that the nonprompt $D^0$ $R_{AA}$ is comparable to $B^+$. The nonprompt $D^0$ $R_{AA}$ is higher than the prompt $D^0$ below 20 GeV/c. These evidences suggest that bottom quarks lose less energy than charm quarks in the QGP medium. Comparing our experimental results with theoretical predictions, we find that the nonprompt $D^0$ is compatible with theories including both collisional and radiative energy losses, for instance, CUJET \cite{CUJET}. Models with only collisional energy loss predict differently at high $p_T$. For example, the PHSD \cite{PHSD} model predicts a much larger suppression than our experimental data at $p_T >$ 10 GeV/c.

\section{$B^+$ and $B^0_s$ mesons}


Both $B^+$ and $B^0_s$ are reconstructed with nonprompt $J/\psi$ and kaons with approximately 500 $\mu$m decay lengths. $B^+$ mesons are reconstructed from the decay channel $B^+ \rightarrow J/\psi K^+ \rightarrow \mu^+ \mu^- K^+$. $B^0_s$ is reconstructed from the decay channel $B^0_s \rightarrow J/\psi \phi \rightarrow \mu^+ \mu^- K^+ K^-$. $B^+$ and $B^0_s$ are selected from the statistically enriched and dedicated dimuon triggered samples. Thanks to the excellent CMS vertexing and tracking performance, B mesons can be reconstructed without hadronic particle identification.

\begin{figure}[hbtp]
\begin{center}
\includegraphics[width=0.32\textwidth]{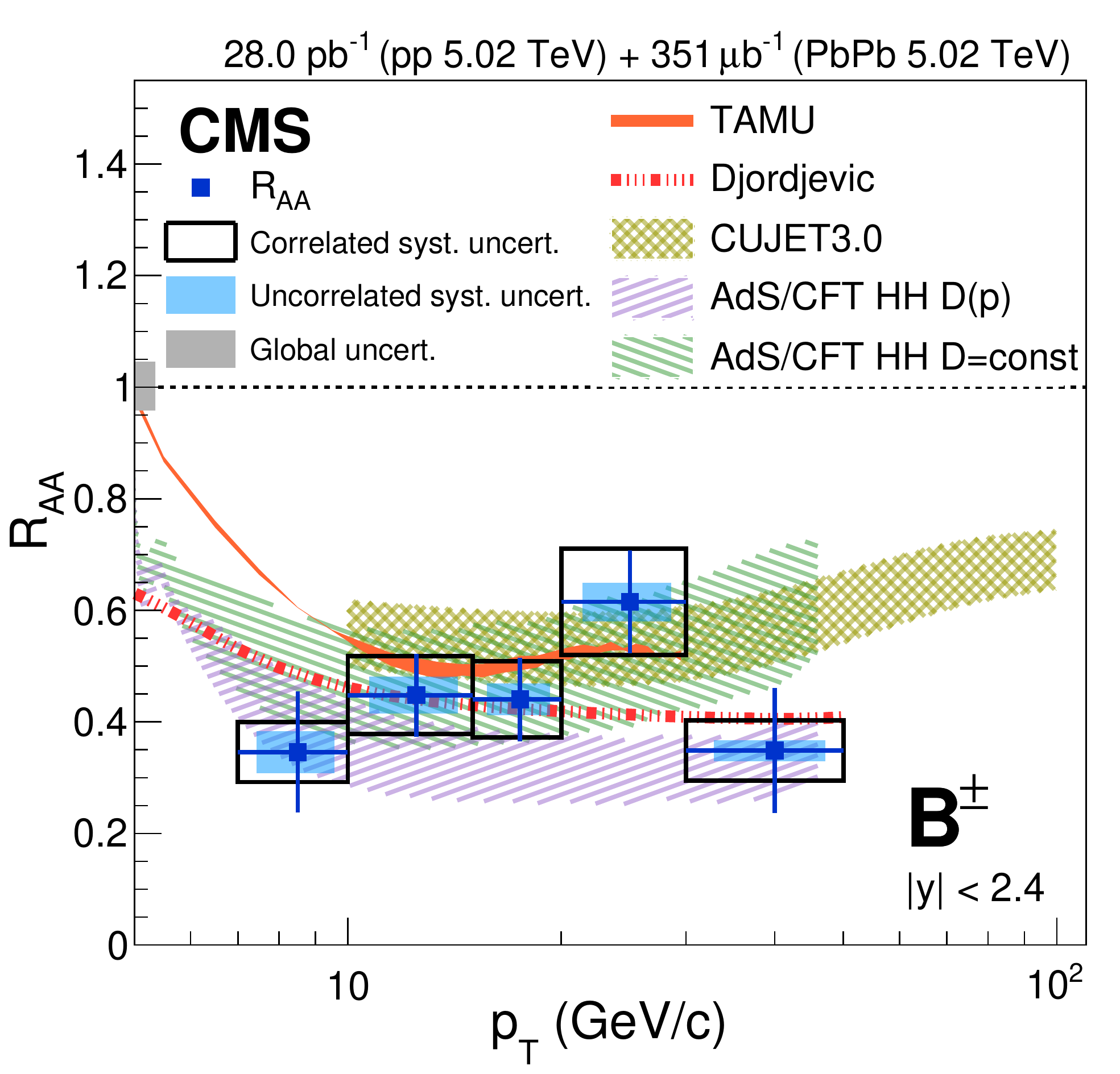}
\includegraphics[width=0.32\textwidth]{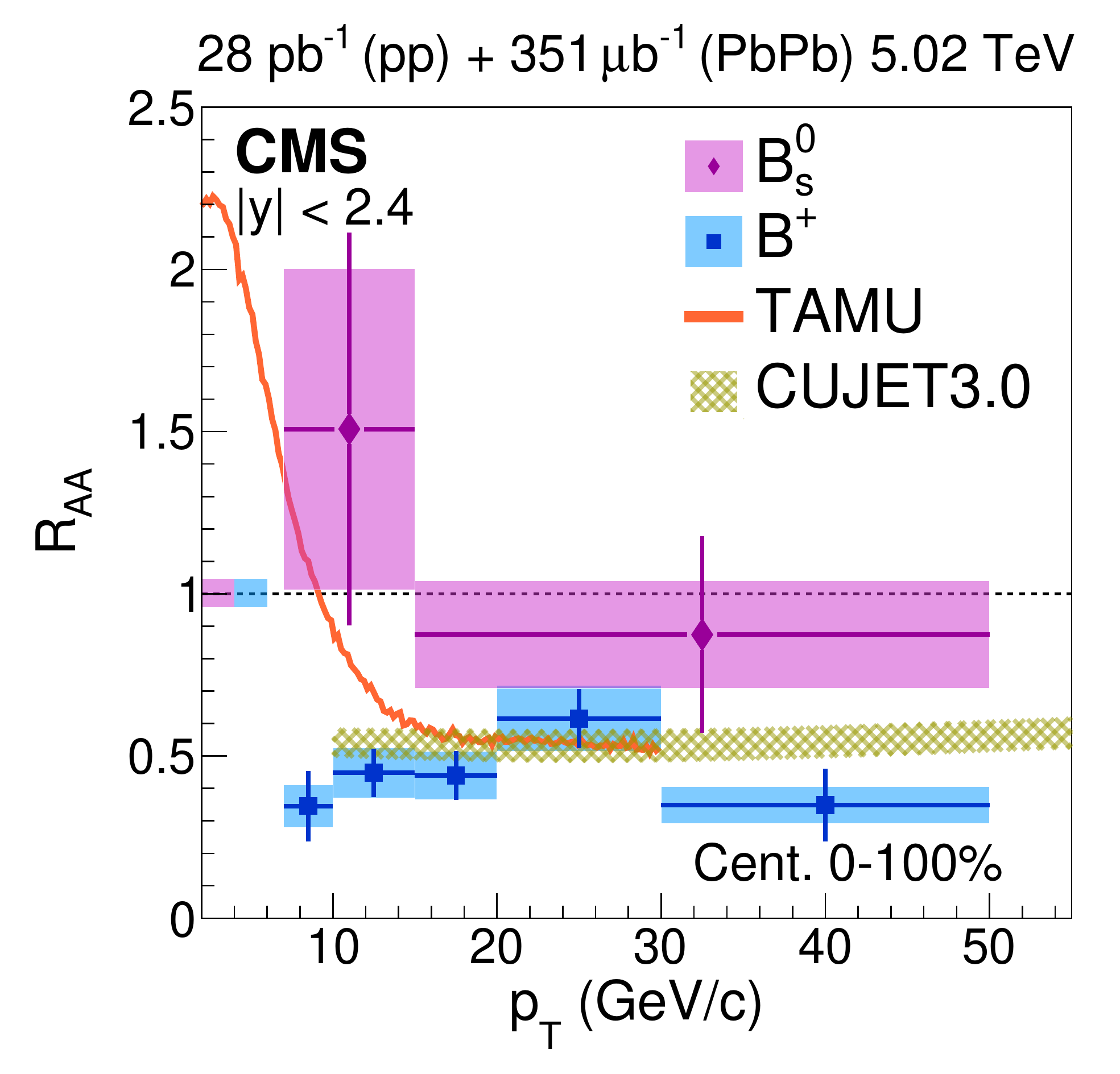}
\includegraphics[width=0.32\textwidth]{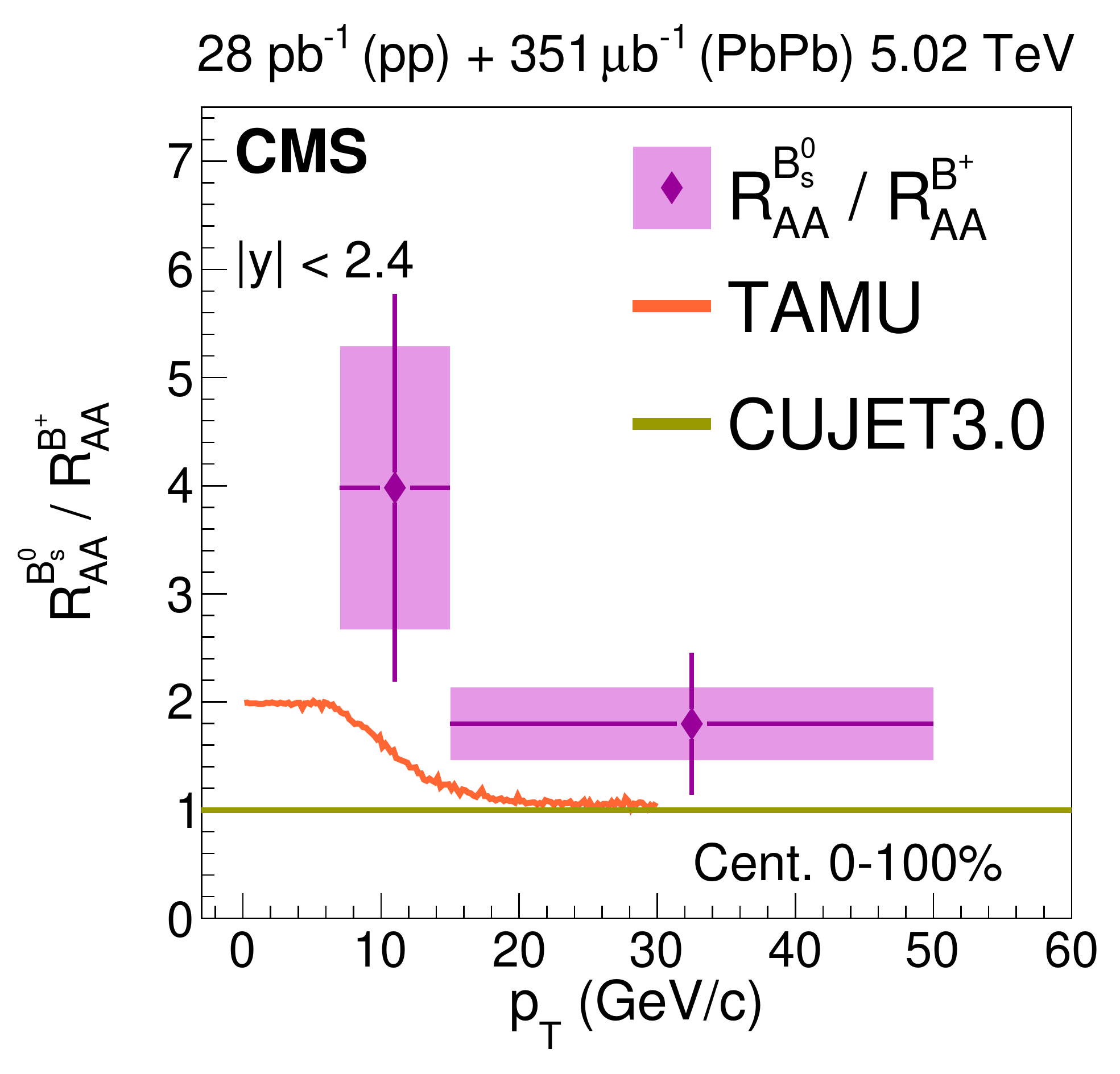}
\caption{Left panel: $B^+$ $R_{AA}$ as a function of $p_T$ \cite{BPlusPaper} and comparison to several theoretical models \cite{CUJET, D0TAMU, Djordjevic, AdSCFT1, AdSCFT2}. Middle panel: $B^0_s$ $R_{AA}$ as a function of $p_{T}$ and comparison with TAMU and CUJET 3.0 \cite{BsPaper}. Right panel: $B^0_s$ to $B^+$ $R_{AA}$ ratio as a function of $p_{T}$ and comparison with TAMU and CUJET 3.0 \cite{BsPaper}.}
\label{fig:BsBPFinalResults}
\end{center}
\end{figure}

The $B^+$ and $B^0_s$ $R_{AA}$ a function of $p_T$, compared with theoretical predictions, are shown on Fig. \ref{fig:BsBPFinalResults}.

A suppression of $B^+$ meson production in PbPb is observed, suggesting that bottom quarks lose energy in the QGP. In addition, there is no significant $p_T$ dependence in the range of 7 -- 50 GeV/c. The $B^+$ $R_{AA}$ is consistent with most theoretical predictions in the range of 7 -- 50 GeV/c. There is a hint for less suppression for $B^0_s$ mesons than $B^+$ mesons, which could be due to strangeness enhancement in the QGP medium. The two measurements are nonetheless compatible given the current size of the uncertainties. Taking the ratio of the $B^0_s$ to $B^+$ $R_{AA}$ allows for the cancellation of correlated systematic uncertainties. There is a hint that $B^0_s$/$B^+$ $R_{AA}$ ratio is larger than unity, as shown in Fig. \ref{fig:BsBPFinalResults}. 

\section{Prompt $D_s^{+}$ mesons}

\begin{figure}[hbtp]
\begin{center}
\includegraphics[width=0.33\textwidth]{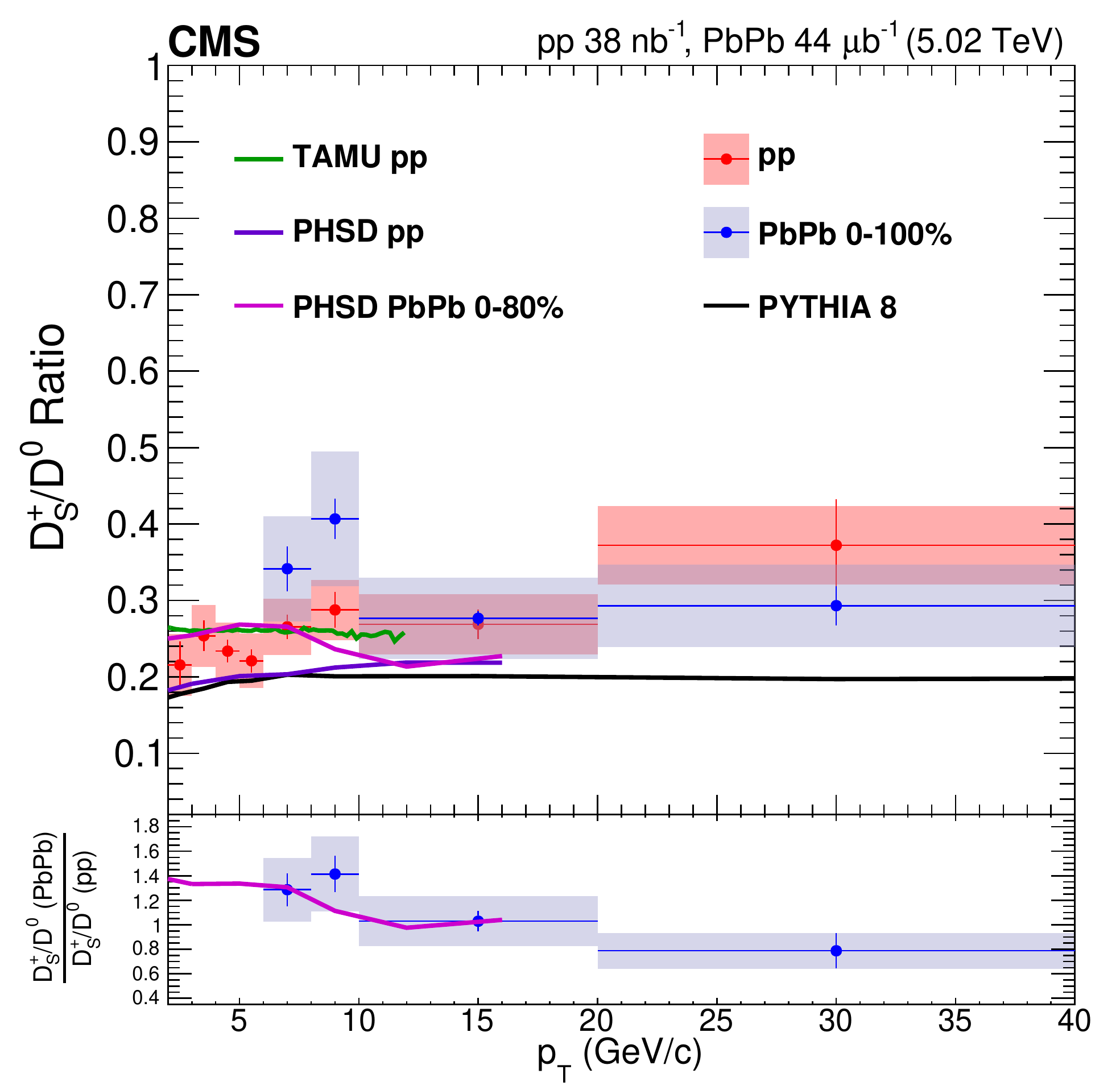}
\caption{The upper panel shows the $D^+_s/D^0$ ratios in pp and PbPb collisions (top) and their ratio (bottom) \cite{DsPaper}. Comparison with theoretical predictions \cite{TAMUDsD0, PHSDDsD0, PYTHIA8} are also presented.} 
\label{fig:D0Ds}
\end{center}
\end{figure}

The $R_{AA}$ of prompt $D^+_s$ mesons, as well as the $D^+_s$ to $D^0$ ratios in pp and PbPb collisions \cite{DsPaper}, can help us better understand the energy loss and hadronization mechanism in the charm sector. The prompt $D^0$ mesons are reconstructed in the $D^0 \rightarrow K^- \pi^+$ channel and the prompt $D^+_s$ are reconstructed in the $D^+_s \rightarrow \phi \pi^+ \rightarrow K^+ K^- \pi^+$ channel. They are presented in Fig. \ref{fig:D0Ds}.

As shown in Fig. \ref{fig:D0Ds}, within uncertainties, no significant strangeness enhancement is found in the $p_{T}$ range from 6 to 40 GeV/c in PbPb compared to pp. Moreover, both TAMU \cite{TAMUDsD0} and PHSD \cite{PHSDDsD0} calculations agree reasonably well with $D^+_s/D^0$ in pp. A good agreement is found between the PHSD model \cite{PHSDDsD0} and $D^+_s/D^0$ double ratio in the data.

\section{Summary}

We have reported the CMS measurements of nonprompt $J/\psi$, nonprompt $D^0$, $B^+$, $B^0_s$, prompt $D^0$, and prompt $D_s^+$ and compared them with theoretical predictions. B mesons and nonprompt D mesons have a smaller suppression than prompt D mesons. Within current uncertainties, we have not observed a significant strangeness enhancement for heavy flavor mesons production associated to the QGP. The larger 2017 (pp) and 2018 (PbPb) datasets are expected to provide more precise and differential measurements in the future.

\section{Acknowledgement}

This work is supported by the United States Department of Energy Nuclear Physics Program and the National Science Foundation Graduate Research Fellowship Program. We would like to thank the Quark Matter 2019 Conference organizers for giving us an opportunity to present this work.

\label{}





\bibliographystyle{elsarticle-num}
\bibliography{<your-bib-database>}







\end{document}